# ADVANCED TECHNOLOGY LARGE-APERTURE SPACE TELESCOPE (ATLAST): A TECHNOLOGY ROADMAP FOR THE NEXT DECADE

**R.F.I. Submitted to NRC ASTRO-2010 Survey**

**Dr. Marc Postman**
**Space Telescope Science Institute**
**Email: postman@stsci.edu**
**Phone: 410-338-4340**

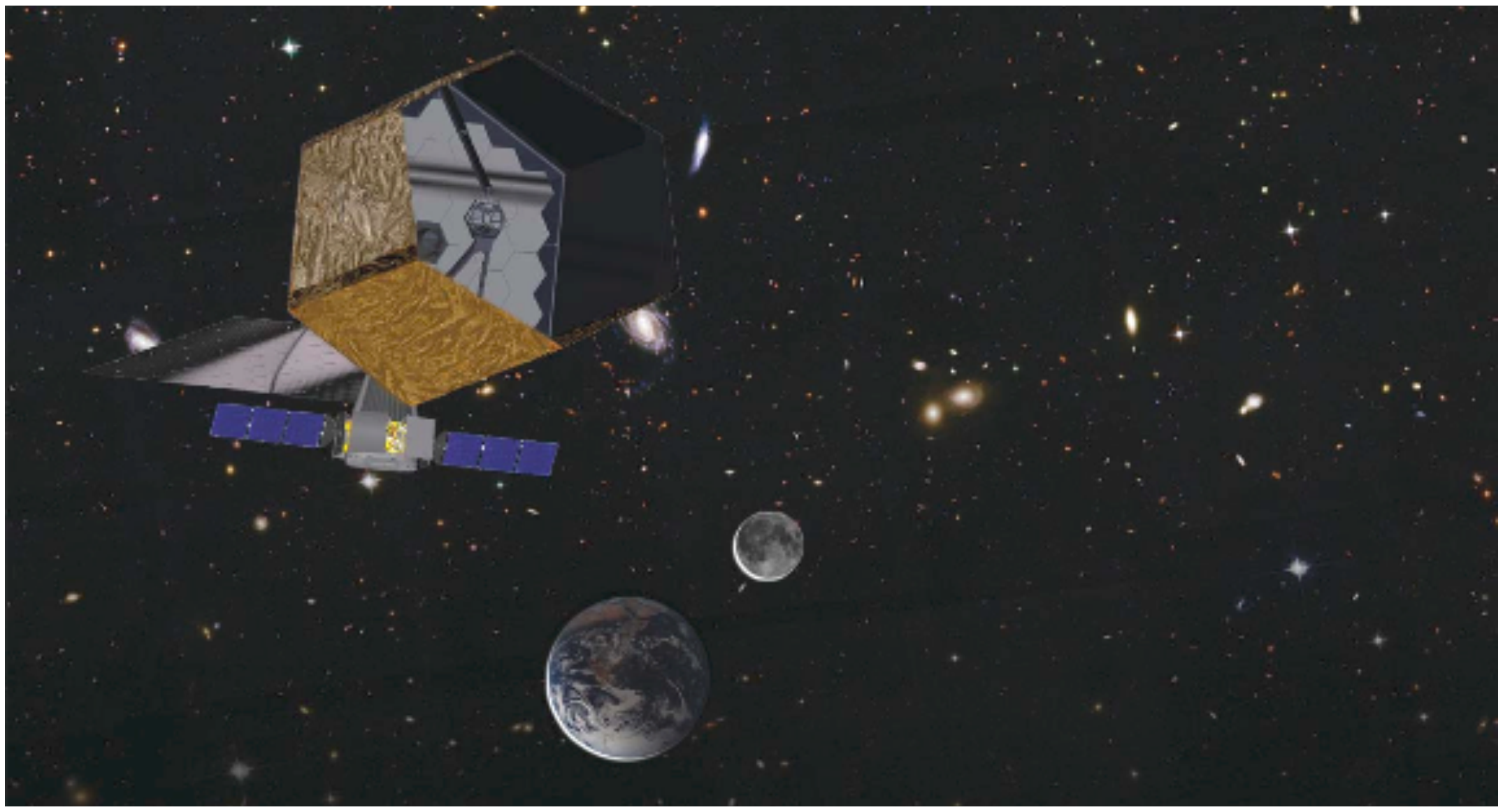

## Co-Investigators

Vic Argabright[1], Bill Arnold[11], David Aronstein[3], Paul Atcheson[1], Morley Blouke[1], Tom Brown[4], Daniela Calzetti[5], Webster Cash[6], Mark Clampin[3], Dave Content[3], Dean Dailey[7], Rolf Danner[7], Rodger Doxsey[4], Dennis Ebbets[1], Peter Eisenhardt[8], Lee Feinberg[3], Andrew Fruchter[4], Mauro Giavalisco[5], Tiffany Glassman[7], Qian Gong[3], James Green[6], John Grunsfeld[9], Ted Gull[3], Greg Hickey[8], Randall Hopkins[2], John Hraba[2], Tupper Hyde[3], Ian Jordan[4], Jeremy Kasdin[10], Steve Kendrick[1], Steve Kilston[1], Anton Koekemoer[4], Bob Korechoff[8], John Krist[8], John Mather[3], Chuck Lillie[7], Amy Lo[7], Rick Lyon[3], Peter McCullough[4], Gary Mosier[3], Matt Mountain[4], Bill Oegerle[3], Bert Pasquale[3], Lloyd Purves[3], Cecelia Penera[7], Ron Polidan[7], Dave Redding[8], Kailash Sahu[4], Babak Saif[4], Ken Sembach[4], Mike Shull[6], Scott Smith[2], George Sonneborn[3], David Spergel[10], Phil Stahl[2], Karl Stapelfeldt[8], Harley Thronson[3], Gary Thronton[2], Jackie Townsend[3], Wesley Traub[8], Steve Unwin[8], Jeff Valenti[4], Robert Vanderbei[10], Michael Werner[8], Richard Wesenberg[3], Jennifer Wiseman[3], Bruce Woodgate[3]

**AFFILIATION CODES:**

**1 = Ball Aerospace & Technologies Corp.**
**2 = Marshall Space Flight Center**
**3 = Goddard Space Flight Center**
**4 = Space Telescope Science Institute**
**5 = Univ. Massachusetts, Amherst**
**6 = Univ. Colorado, Boulder**
**7 = Northrop Grumman Aerospace Systems**
**8 = Jet Propulsion Laboratory, California Institute of Technology**
**9 = Johnson Space Flight Center**
**10 = Princeton Univ.**
**11 = Jacobs ESTS Group @ MSFC**

## Executive Summary

For four centuries new technology and telescopes of increasing diameter have driven astronomical discovery for the simple reason that astronomy is a photon-limited field. The *Hubble Space Telescope* (*HST*), to date the largest UV/optical astronomical space telescope, has demonstrated the breadth of fundamental astrophysics that can be extracted from space-based observations in the UV-optical-near IR. *HST's* versatility has allowed it to be used to make pioneering discoveries in fields never envisioned by its builders. The paradigm-shifting discoveries in the next two decades will be made with ever more capable instruments and facilities. Here we outline the technology developments and mission concepts required for the next step – a highly versatile UV-optical-near IR observatory in space, larger and more capable than either *HST* or its IR-optimized successor, the *James Webb Space Telescope* (*JWST*). Although substantial investments are required for the next steps, the basic technologies needed either already exist or we understand the path forward, allowing us to construct schedules, budgets, and the main decision points.

The **Advanced Technology Large-Aperture Space Telescope** (*ATLAST)* is a set of mission concepts for the next generation of UVOIR space observatory with a primary aperture diameter in the 8-m to 16-m range that will allow us to perform some of the most challenging observations to answer some of our most compelling questions, including "Is there life elsewhere in the Galaxy?" We have identified two different telescope architectures, but with similar optical designs, that span the range in viable technologies. The architectures are a telescope with a monolithic primary mirror and two variations of a telescope with a large segmented primary mirror. This approach provides us with several pathways to realizing the mission, which will be narrowed to one as our technology development progresses. The concepts invoke heritage from *HST* and *JWST* design, but also take significant departures from these designs to minimize complexity, mass, or both.

We have a plan to develop the needed technologies over the course of 6 to 9 years with a budget of $287M. Concurrently with the technology investment, a serious mission concept competition should be held to develop and select the concept and the prime contractor through Phase A, following the model of the *JWST* project. Such studies would cost about $48M. With these investments, and the support of NASA and the worldwide astronomical community, *ATLAST* could be ready to start its Phase B as early as 2018, leading to a launch 7 to 10 years later.

Our report provides details on the mission concepts, shows the extraordinary scientific progress they would enable, and describes the most important technology development items. These are the mirrors, the detectors, and the high-contrast imaging technologies, whether internal to the observatory, or using an external occulter. Experience with *JWST* has shown that determined competitors, motivated by the development contracts and flight opportunities of the new observatory, are capable of achieving huge advances in technical and operational performance while keeping construction costs on the same scale as prior great observatories.

***We urge the Decadal Survey to endorse the serious investments in technology and mission concept development that would enable the Advanced Technology Large Aperture Space Telescope (ATLAST) to take its place with HST and JWST in opening the Universe to human knowledge and exploration.***

## 1. Scientific Motivations for *ATLAST*

Conceptual breakthroughs in understanding astrophysical phenomena happen when our observatories allow us to detect and characterize faint structure and spectral features on the relevant angular scales. By virtue of its ~12 milli-arcsecond angular resolution at ~500 nm coupled with its ultra high sensitivity, superb stability and low sky background, *ATLAST* will make these breakthroughs – both on its own and in combination with other telescopes with different capabilities. *ATLAST* has the performance required to detect the potentially rare occurrence of biosignatures in the spectra of terrestrial exoplanets, to reveal the underlying physics that drives star formation, and to trace the complex interactions between dark matter, galaxies, and the intergalactic medium. Because of the large leap in observing capabilities that *ATLAST* will provide, we cannot fully anticipate the diversity or direction of the investigations that will dominate its use – just as the creators of *HST* did not foresee its pioneering roles in characterizing the atmospheres of Jupiter-mass exoplanets or measuring the acceleration of cosmic expansion using distant supernovae. It is, thus, essential to ensure *ATLAST* has the versatility to far outlast the vision of current-day astronomers. We discuss briefly a small subset of the key scientific motivations for *ATLAST* that we can conceive of today.

### Does Life Exist Elsewhere in the Galaxy?

We are at the brink of answering two paradigm-changing questions: *Do other planets like Earth exist? Do any of them harbor life?* The tools for answering the first question already exist (e.g., *Kepler*); those that can address the second can be developed within the next 10-20 years [1]. ***ATLAST* is our best option for an extrasolar life-finding facility** and is consistent with the long-range strategy for space-based initiatives recommended by the Exoplanet Task Force [2]. *ATLAST* has the angular resolution and sensitivity to characterize the atmosphere and surface of an Earth-sized exoplanet in the Habitable Zone (HZ) at distances up to ~45 pc, including its rotation rate, climate, and habitability. These expectations are based on our simulated exoplanet observing programs using the known stars, space telescopes with aperture diameters ranging from 2 m to 16 m, and realistic sets of instrumental performance parameters and background levels.

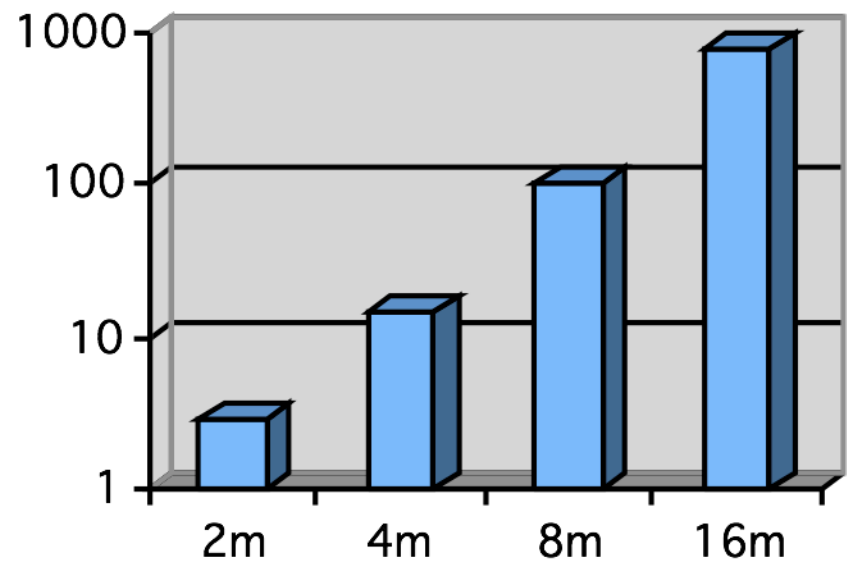

***Figure. 1:*** The average number of F,G,K stars where SNR=10 R=70 spectrum of an Earth-twin ***could*** be obtained in < 500 ksec as a function of telescope aperture, D. The growth in the sample size scales as $D^3$.

We start by selecting spectral type F,G,K stars from the Hipparcos catalog and identify, for each telescope aperture, D, those stars whose HZ exceeds an inner working angle (IWA) of 3λ/D at 760 nm (the $O_2$ absorption feature). As a detection goal, we assume that each star has an Earth-twin in its HZ (with Δmag = 25), realizing that super Earths will be easier targets. We include plausible instrumental efficiencies and noise properties, and assume a 3-zodi background (local plus exosolar). We assume that our starlight suppression system (either internal coronagraph or external occulter) is capable of achieving a suppression of at least 25 mags ($10^{-10}$) and include residual background from the star as an additional noise source. We then identify the number of stars for which an R=70 spectrum with signal-to-noise ratio (SNR) of 10 at 760 nm could be acquired in 500 ksec or less. The results, averaged over different simulations done using various starlight suppression options (internal coronagraphs of various kinds as well as an external occulter), are shown in Figure 1. To estimate the number of potentially inhabited worlds detected, one must multiply the numbers in Figure 1 by the fraction of the FGK stars that have an Earth-sized planet in their HZ ($\eta_{\oplus}$) and also by the fraction of those exo-Earths that have detectable biosignatures. The values of these fractions are currently not constrained but their product is not likely to be close to unity. One must conclude that **to maximize the chance for a successful search for life in the solar neighborhood requires a space telescope with an aperture size of *at least* 8 meters**.

Estimates of the SNR of habitability and biosignature features in an Earth-twin spectrum, achievable with *ATLAST*, are shown in Table 1. For these calculations we use a fully validated model of the Earth's spectrum [3,4], in combination with the observed visible reflection spectrum of the present Earth. We assume that the exoplanet is at maximum elongation and that the planet is observed for a length of time sufficient to achieve an SNR of 10, at a spectral resolution R = 70, in the red continuum. The Rayleigh (air column) signal is the blue enhanced albedo from atmospheric molecules. The $O_3$ and $O_2$ signals are biosignatures. The cloud/surface signal around 750 nm will vary with time as the planet rotates, and is therefore a rotation signature. The vegetation signal is the enhanced albedo of the Earth, from land plants, for wavelengths longer than ~720 nm [5], with a modest SNR. The $H_2O$ signal is a prime habitability indicator. Column 3 gives the width of the spectral feature. All of these SNR values can easily be improved with re-visits. In addition, *ATLAST* will allow us to glean substantial information about an exo-Earth from temporal variations in its features. Such variations inform us about the nature of the dominant surface features, changes in climate, changes in cloud cover (weather), and, potentially, seasonal variations in surface vegetation [6]. Constraints on variability require multiple visits to each target. **The 8-m *ATLAST* (with internal coronagraph) will be able to observe ~100 different star systems 3 times each in a 5-year interval and not exceed 20% of the total observing time available to the community. The 16-m version (with internal coronagraph) could visit up to ~250 different stars 3 times each in a 5-year period. The 8-m or 16-m *ATLAST* (with a single external occulter) can observe ~85 stars 3 times each in a 5-year period, limited by the transit times of the occulter.** Employing multiple occulters would remove this limitation.

***Table 1:* Habitability and Bio-Signature Characteristics**

| Feature | λ (nm) | Δλ (nm) | SNR | Significance |
|---|---|---|---|---|
| Reference continuum | ~750 | 11 | 10 | |
| Air column | 500 | 100 | 4 | Protective atmosphere |
| Ozone ($O_3$) | 580 | 100 | 5 | Source is oxygen; UV shield |
| Oxygen ($O_2$) | 760 | 11 | 5 | Plants produce, animals breathe |
| Cloud/surface reflection | 750 | 100 | 30 | Rotation signature |
| Land plant reflection | 770 | 100 | 2 | Vegetated land area |
| Water vapor ($H_2O$) | 940 | 60 | 16 | Needed for life |

With *ATLAST,* we will be able to determine if HZ exoplanets are indeed habitable, and if they show signs of life as evidenced by the presence of oxygen, water, and ozone. *ATLAST* also will provide useful information on the column abundance of the atmosphere, the presence of continents and oceans, the rotation period, and the degree of daily large-scale weather variations.

## Exploration of the Modern Universe

We know that galaxies form and evolve but we know little about how this happens. The physical processes involve complex interactions between the baryonic matter in the galaxies, the energy exchanged during the birth and death of stars, the gas outside the galaxies in the intergalactic medium (IGM), other neighboring galaxies, and the dark matter that dominates and shapes the underlying gravitational potential. Revealing the physics behind galaxy formation and evolution requires making a broad array of observations from the current epoch to the epoch of the first stars. *ATLAST* will provide many of the key pieces needed to solve this puzzle, particularly in the redshift range $z < 4$ when the cosmic star formation rate peaks and then fades, and galaxies develop their current morphologies.

***ATLAST* will enable extensive probes of the local IGM in the UV, revealing the nature of its interaction with galaxies.** Understanding how gas in the IGM gets into galaxies and how galaxies respond to inflow lies at the heart of understanding galactic evolution. The mode of accretion depends on the depth of the potential well (galaxy type) and the location at which the intergalactic gas is shocked as it encounters that potential [7,8]. Depending on the mass of the galaxy halo, the infalling gas may be shocked and heated or accrete in "cold mode" along narrow filaments. Gas can also be removed from galaxies via tidal and ram pressure stripping, or during the accretion of gas-rich dwarfs onto giant galaxies. Metal-enriched gas introduced into the IGM by these processes will be dynamically cool. All of these accretion and gas removal theories have observational consequences (e.g., Fig. 2) that can be tested if the distribution of gas in the cosmic web around galaxies can be characterized through absorption and emission line spectroscopy. The observational challenge is to acquire datasets of sufficient sample size and with enough diagnostic power **(i.e., spectral resolution)** to identify and characterize the various processes at work. The number of suitable background sources available for absorption line measurements is presently limited by the sensitivity of current instrumentation (*HST*/*COS*).

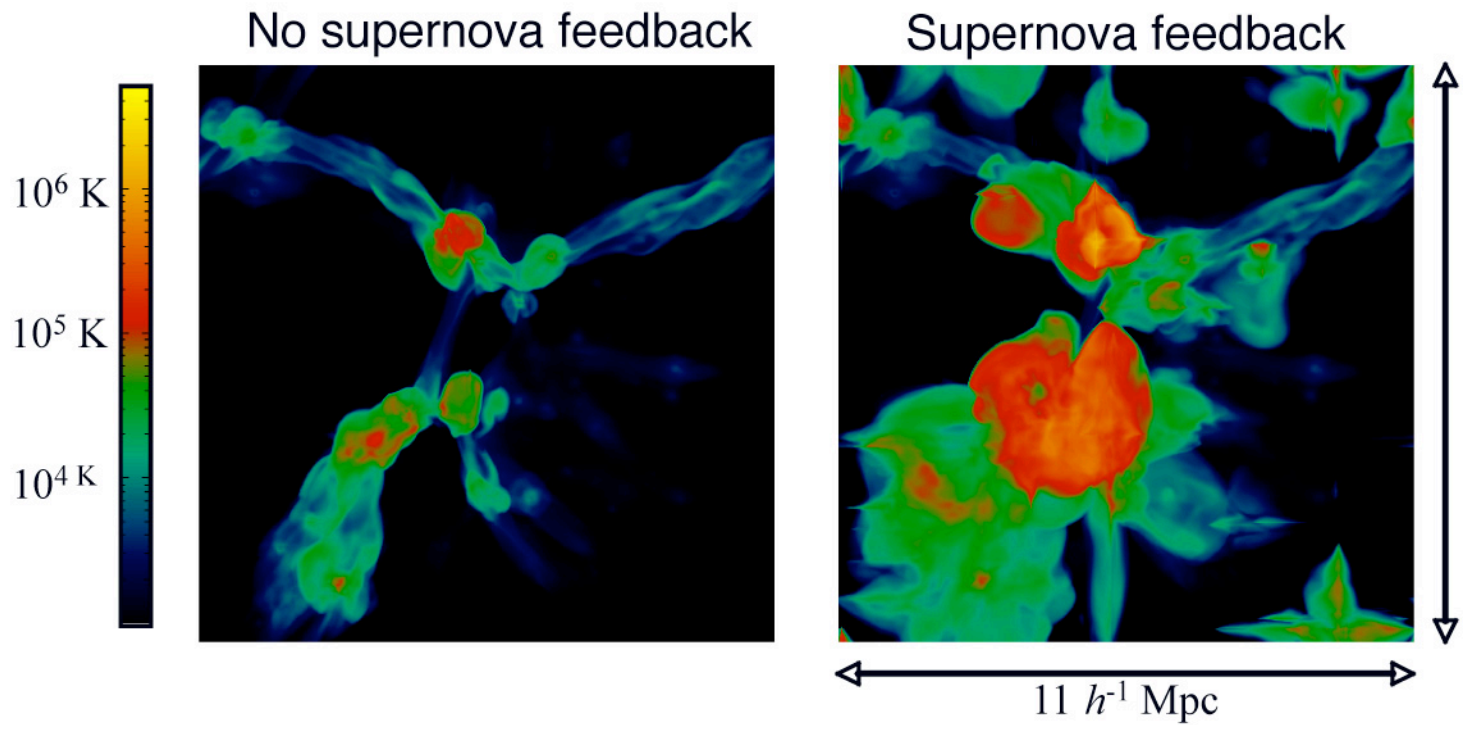

***Figure 2*:** IGM gas temperature distribution for cosmological models with and without supernova feedback [9].

*ATLAST's* dramatically increased absorption line sensitivity at UV and optical wavelengths is crucial for reaching the required background source densities. There are ~100 quasars per square degree that are brighter than a *GALEX* flux of $m_{FUV}$ ~ 24 [10]. At this sampling, one can select sight lines next to thousands of examples of any common galaxy, group, or cluster. *ATLAST* could then be used to produce a high-resolution map of the gas and metals surrounding these structures, which could be used to compare directly against simulation predictions [11,12]. With *ATLAST* one could also use multiple quasars *and* distant galaxies as background continuum sources to dissect the gas distribution in fields known to have galaxies and gas at the same redshift [13]. *ATLAST*'s large aperture will enable contiguous regions of ~10 Mpc on a side (like in Fig. 2) to be surveyed in about 2 weeks of exposure time (with an 8-m *ATLAST* with enhanced UV detectors) or in ~3 days (with a 16-m *ATLAST* with enhanced UV detectors) [14]. *ATLAST* could also be used systematically to target individual nearby galactic coronae and groups of galaxies, for which it would be possible to observe the production sites of heavy elements (star-forming regions, SNe, emission nebulae), follow the processes by which the elements are transported within galaxies, and trace the expulsion of metals into the IGM.

Determining whether the stellar initial mass function (IMF) is universal or environment-dependent places further stringent constraints on the evolution of the baryonic component of galaxies. *ATLAST* will uniquely be able to measure the IMF down to 1 $M_{SUN}$ out to 10 Mpc over the full range of star forming environments (which are outside the range of *HST* and *JWST*), including environments similar to those found in high redshift galaxies. This will yield fundamental predictions for a comprehensive theory of star formation [15].

***ATLAST* will, for the first time, enable star formation histories to be reconstructed for hundreds of galaxies beyond the Local Group, opening the full range of star formation environments to exploration.** A comprehensive and predictive theory of galaxy formation and evolution requires that we accurately determine how and when galaxies assemble their stellar populations, and how this assembly varies with environment. By definition the dwarf galaxies we see today are not the same as the dwarf galaxies and proto-galaxies that were disrupted during assembly. Our only insight into those disrupted building blocks comes from sifting through the resolved field populations of the surviving giant galaxies to reconstruct the star formation history, chemical evolution, and kinematics of their various structures [16]. **Resolved stellar populations are cosmic clocks.** Their most direct and accurate age diagnostic comes from resolving both the dwarf and giant stars, including the main sequence turnoff. But the main sequence turnoff rapidly becomes too faint to detect with any existing telescope for any galaxy beyond the Local Group. This greatly limits our ability to infer much about the details of galactic assembly because the galaxies in the Local Group are not representative of the galaxy population at large. *ATLAST* will allow us to reach well beyond the Local Group as shown in Figure 3. *HST* and *JWST* cannot reach any large galaxies besides our Milky Way and M31 (see Figure 3). An 8-meter space telescope can reach 140 galaxies including 12 giant spirals and the nearest giant elliptical. A 16-meter space telescope extends our reach to the Coma Sculptor Cloud, netting a total of 370 galaxies including 45 giant spirals and 6 ellipticals. Deriving ages and other galactic properties from color-magnitude data requires photometry for thousands of stars spanning 4 orders of magnitude in luminosity. **This implies the need for a wide-field imager on *ATLAST* with half-Nyquist sampling over a field of view of at least 4 arcminutes.** *ATLAST* will work in concert with 30m-class ground-based telescopes (e.g., *TMT*), expanding our reach to other well-populated galaxy groups, with *ATLAST* obtaining photometry of V~35 magnitude G dwarf stars and *TMT* obtaining kinematics of much brighter giants out to the Coma Sculptor Cloud. The dwarf stars in the Coma Sculptor Cloud are effectively inaccessible to *TMT*, requiring gigaseconds of integration even for an isolated star.

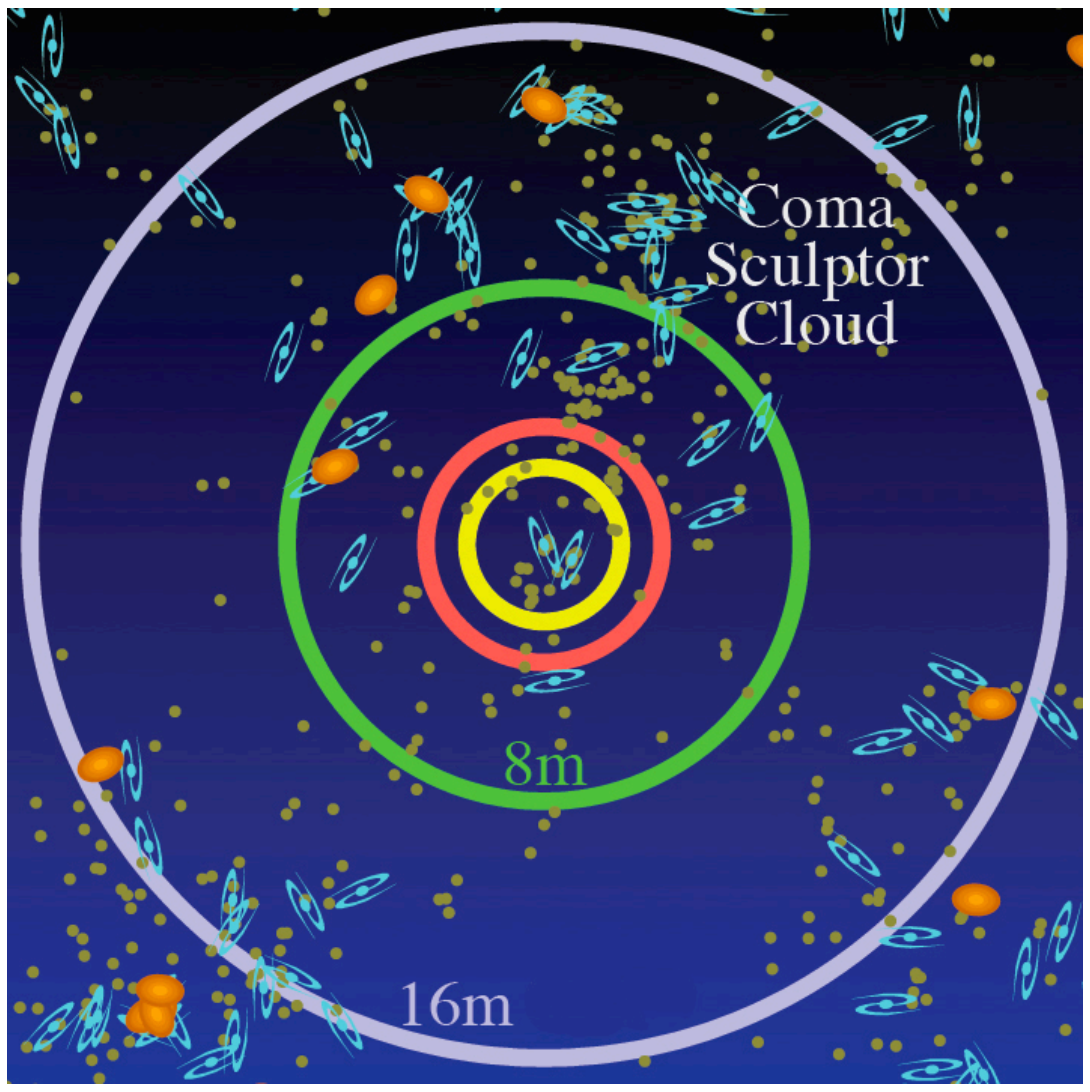

***Figure 3*:** Map of local universe (24 Mpc across) shown with the distances out to which *HST* (yellow), *JWST* (orange), and *ATLAST* (8-m, 16-m), can detect solar analogs in V and I passbands at SNR=5 in 100 hours. Giant spirals, like M31, are indicated by the blue galaxy symbols, giant ellipticals as orange blobs, and dwarf galaxies as small green dots.

**Synergy with Other Astronomical Facilities:** The impressive capabilities anticipated for the 20-m to 40-m optical/near infrared (NIR) ground-based observatories in the upcoming decade will redefine the existing synergy between ground and space telescopes. Adaptive optics (AO) may enable ground-based optical telescopes to achieve intermediate Strehl ratios (~40%-80%) over a field of view of up to 2 arcminutes at wavelengths longer than ~1000 nm [17,18,19,20]. Extreme AO may even enable diffraction-limited performance at ~600 nm in a ~2 arcsecond field of view, albeit with extremely limited sky coverage.

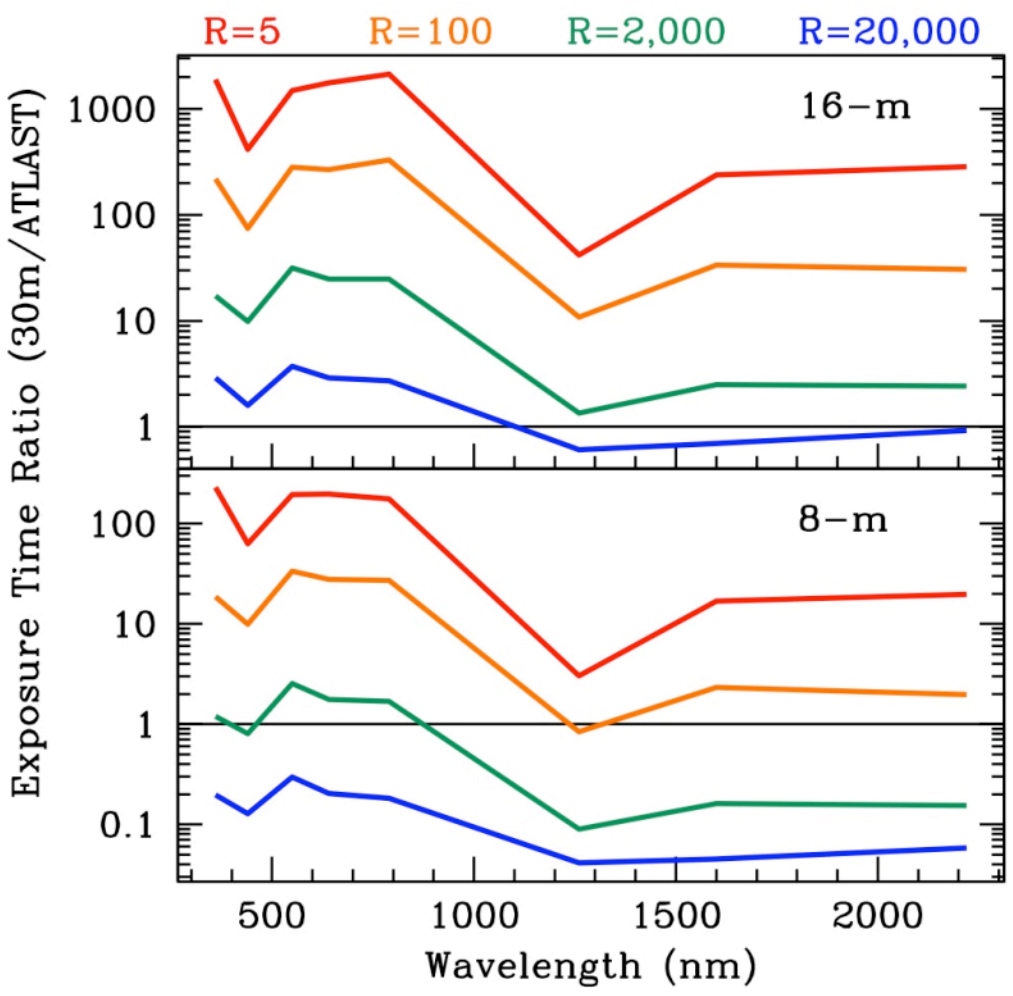

***Figure 4:*** Exposure time ratio to reach SNR=10 between a ground-based 30-m (with AO in the near-IR) and *ATLAST* 8-m and 16-m as a function of spectral resolution, R, and wavelength. *ATLAST* is faster when ratio is greater than 1. The ratios for R=5, 100, 2000, and 20000 are shown as red, orange, green, and blue, respectively.

Space telescopes, of course, provide exclusive access to UV wavelengths. But there are unassailable advantages of space telescopes for observations in the optical and NIR range:

- Ultra-deep panoramic imaging (tens of arcminutes or more)
- High Strehl ratios (>90%) that are highly stable (<2% variations) spatially and temporally
- Very high-contrast imaging ($<10^{-8}$)
- Ultra precise photometry (<0.0001 magnitude) and astrometry (<<1 mas)
- Ability to achieve any of the above on demand (no day/night or weather at L2 orbit)

As one example, in Figure 4 we compare the time to reach a SNR=10 for a point source as a function of spectral resolution, R, for a 30-m ground-based telescope with those for an 8-m and 16-m *ATLAST*. All are assumed to have the same instrument/detector performance. The 30-m telescope is assumed to be diffraction-limited at wavelengths longer than 1000 nm and seeing-limited (0.4 arcsec) below 1000 nm. An 8-m *ATLAST* would be 10 to 100 times faster than a 30-m on the ground for imaging (R=5) and up to 40 times faster for low-res (R=100) spectroscopy. A 16-m space telescope would be up to 3000 times faster than a 30-m on the ground for imaging; 20 to 300 times faster for R=100 spectroscopy from 300 – 2500 nm and typically 10 to 30 times faster for R=2000 optical spectroscopy. Large ground-based telescopes will, in general, be faster for most medium and high-resolution (R > 2000) spectroscopy, at least for sources brighter than ~28 AB magnitude. *ATLAST* imaging would reach the depth of the *HST* Ultra Deep Field ~100 times faster (for 8-m) to >1300 times faster (for 16-m) than *HST* but with higher spatial resolution, enabling panoramic ultra deep fields covering hundreds of times the original UDF area.

*ATLAST*'s high spatial resolution, coupled with its sensitive high-resolution UV/optical spectroscopic capabilities, will make it an ideal companion to *ALMA (*e.g., to fully interpret proto-planetary disk chemistry and dynamics). *ALMA* will achieve 5 milli-arcsecond resolution at its highest frequency, 900 GHz. *ATLAST* will complement in-situ spacecraft by enabling long-term monitoring of the outer planets and small bodies within our Solar System. Its large collecting area will enable UV imaging, below 200 nm, of small bodies beyond Jupiter revealing subtle albedo variations and the possible presence of simple organic molecules.

## 2. *ATLAST* Technical Overview

The *ATLAST* technology development plan is based on three point designs the team developed using funding from NASA's Astrophysics Strategic Mission Concept Study program, NASA/MSFC, NASA/GSFC, and related studies at JPL, Northrop Grumman, and Ball Aerospace. Two of the concepts, the 8-m monolithic mirror telescope (hereafter *ATLAST-8m)* and the 16.8-m segmented mirror telescope (*ATLAST-16m*), span the range of UVOIR observatories that are enabled by NASA's proposed Ares-V launch vehicle. *ATLAST-8m* is studied because of the inherent advantages offered by a monolithic aperture telescope in terms of high-contrast imaging and wavefront error (WFE) control. *ATLAST-16m* is studied because it represents a pathway to truly large apertures in space and uses the largest extrapolation of a *JWST*-like chord-fold primary mirror packaging. However, the *ATLAST* mission is not solely dependent on Ares V. Our third concept, a 9.2-m segmented telescope (*ATLAST-9.2m)*, is compatible with an Evolved Expendable Launch Vehicle (EELV) and also adopts *JWST* design heritage.

All *ATLAST* concepts require many of the same key technologies. **We believe these designs compose a robust set of options for achieving the next generation of UVOIR space observatory in the 2020 era.** Their technologies also will enable a new generation of more capable small and medium class space-based and balloon-borne observatories.

Key to the *ATLAST-8m* and *ATLAST-16m* is NASA's proposed Ares V heavy lift launch vehicle planned for 2019. Ares V (configuration LV 51.00.48) is projected to have the ability to launch an 8.8 m diameter, ~65,000 kg payload into a Sun-Earth Lagrange point (SE-L2) transfer orbit. The current baseline Ares V shroud is a biconic fairing with a 10 m outer diameter and a height of 21.7 m. Ares V will have an 8.8-m diameter by 17.2-m tall dynamic envelope, and a payload volume of 860 cubic meters – nearly three times the volume of the Space Shuttle payload bay. NASA is considering a 'stretch' fairing that would measure 26 m in height, with a volume of 1410 cubic meters.  Finally, a serious trade is underway to replace the biconic nose cone with an ogive-shaped structure. The ogive configuration would provide even more payload volume and useable internal vertical height [21].

**There are several fundamental features common to all our designs.** All *ATLAST* concepts are designed to operate at SE-L2. The optical designs are diffraction limited at 500 nm (36 nm rms WFE) and the optical telescope assembly (OTA) operates near room temperature (280K – 290K). All OTAs employ two, simultaneously usable foci – a three-mirror anastigmat (TMA) channel for multiple, wide field of view (FOV) instruments, and a Cassegrain channel for high-throughput UV instruments and instruments for imaging and spectroscopy of exoplanets (all designs have rms WFE of <5 nm at <2 arcsec radial offset from Cass optical axis).

***Table 2:*** **Tentative *ATLAST* Science and Facility Instruments and their FOV**

| **Telescope** | TMA Focal Plane Instruments | | | | Cass Focal Plane Instruments | | |
|---|---|---|---|---|---|---|---|
| | Vis/NIR Wide-field Imager | Vis/NIR Multi-Object Spectrograph | Vis/NIR IFU | FGS (FOV per FGS unit) | UV IFU & Spectrograph | Starlight Suppression | Exoplanet Imager & Spectrograph |
| **8-m, 9.2-m** | 8x8 arcmin | 4x4 arcmin | 2x2 arcmin | 3x3 arcmin | 30 arcsec | Internal Coronagraph or Starshade Sensor | ~10 arcsec |
| **16.8-m** | 4x4 arcmin | 3x3 arcmin | 1x1 arcmin | ~1x3 arcmin | 15 arcsec | | ~10 arcsec |

All instruments are modular and could be replaceable on orbit. Fine guidance sensors (FGS) are deployed on the TMA focal plane. Nominal designs incorporate three to four FGS units. Table 2 summarizes potential instrument suites for *ATLAST*. The WF visible/NIR imager employs a beam splitter to simultaneously send light to visible CCD (450-1000 nm) and NIR HgCdTe (1-2.5 microns) detector arrays. The CCD array requires ~1 gigapixels to permit half-Nyquist (*a.k.a.* critical) sampling at 500 nm.

Finally, all *ATLAST* concepts have propellant loads sized to provide ten years of on orbit station keeping and momentum unloading. *ATLAST* can be designed to enable on-orbit servicing to extend the mission lifetime via autonomous rendezvous and docking technology, which Orbital Express recently demonstrated. However, *ATLAST* can achieve high scientific impact without servicing. We summarize the *ATLAST* design concepts below and in Table 3.

***Table 3:*** **Summary of *ATLAST* Point Designs**

| Aperture (meters) | Wavelength Coverage | Orbit | Primary Mirror | Secondary Mirror | Pointing (mas) | Launch Vehicle | Total Mass (kg) | Total Power (kW) |
|---|---|---|---|---|---|---|---|---|
| **8.0** | 110 – 2500 nm | SE-L2 Halo Orbit | Monolithic | On-axis or Off-axis | 1.6 | Ares V | ~59,000 | 11 |
| **9.2** | | | Segmented | On-axis | 1.4 | EELV | ~15,700 | 5.7 |
| **16.8** | | | Segmented | On-axis | 0.8 | Ares V | ~30,000 | ~10 |

(Total Mass and Total Power values include at least a 28% contingency)

## 8-meter Monolithic Mirror Telescope

The *ATLAST-8m* mission concept takes advantage of the volume capacity of the Ares V's fairing to launch an 8-meter monolithic mirror. The Ares V's enormous mass capacity significantly reduces cost and risk by allowing a simple high-mass margin high-TRL design approach. *ATLAST-8m* could be flown early in the 2020 decade.

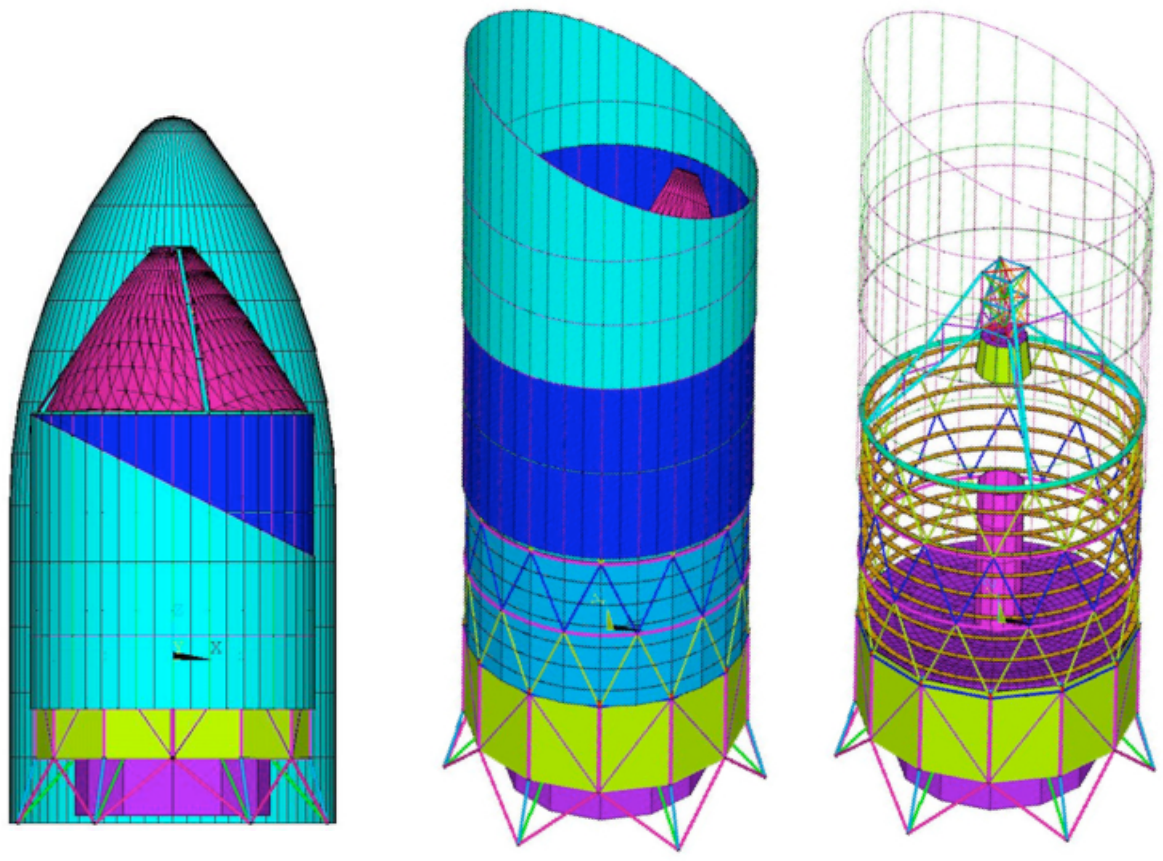

***Figure. 5:*** (Left) 8-m *ATLAST* in Ares V fairing. (Center) After deployment of the sunshield. (Right) Cutaway view.

*ATLAST-8m* uses a modified *HST*-style optical bench for stable primary to secondary mirror alignment. Behind the primary mirror is a 4 m diameter by 4.5 m long bay in which modular science instruments have direct access to the Cassegrain focus and **two** off-axis wide fields of view. A *Kepler*-style 60-degree scarfed sunshield provides thermal isolation, stray-light rejection and environmental protection (see Fig. 5). The sunshade stabilizes the telescope at 135K and active zonal heating raises its temperature to 280K ± 0.1K. The primary mirror's uniform CTE, mass and thermal capacity provides a thermal figure stability of 1 nm with a 500-hour thermal time constant. *ATLAST-8m* offers exceptional thermal stability for long-duration observations regardless of slew or roll angle.

The single most important element of *ATLAST*-8m is the primary mirror. While a *HST*-style lightweight mirror could be used with a *JWST*-class design-margin structure, the Ares V's mass capacity allows us to reduce risk and cost by using a solid meniscus glass mirror – the kind typically used for ground based telescopes (e.g., *Gemini*, *VLT*). In fact we propose to space qualify and launch the existing *VLT* spare or a new off-axis mirror blank. **This approach saves cost because the handling infrastructure exists and the ability to fabricate the mirror to 8 nm rms has been demonstrated.** Facilities to fabricate lightweight 4-m to 8-m class mirrors do not exist. Additionally, *ATLAST-8m* uses large design margins, which reduce cost and risk. An iso-grid truss-structure support maintains primary mirror launch loads below 600 psi, which is well below the 7000 psi design limit for polished glass meniscus substrates. By comparison, lightweight mirrors have a design limit of only 1000 psi.

The entire observatory (telescope, science instruments and spacecraft) has an estimated dry mass of ~51,400 kg with a 45% margin against the Ares V 65,000 kg capacity (excluding the primary mirror whose mass can only decrease as material is removed). If the Ares V mass capacity were to decrease, our design mass can be reduced to maintain a minimum mass margin of 30%. The spacecraft reaction wheels are sized to provide a maximum slew rate of 0.67 deg per minute, roll rate of 0.5 deg per minute and a maximum uninterrupted observation period of 75 hours between momentum dumps. This observation time is assisted by using the 72 $m^2$ of solar panels, mounted behind the observatory on a rotary arm, as a kite tail to balance the solar radiation pressure. Body pointing the telescope using active isolation between the spacecraft and telescope achieves fine pointing stability of 1.6 mas. Four *JWST*-style 'staring' FGS units (two in each corner of the two TMA foci) provide pointing control, roll control and redundancy.

We also examined an off-axis monolithic telescope concept (to optimize the performance of an internal coronagraph for exoplanet observations). A 6 x 8 meter off-axis elliptical mirror is feasible in an Ares V fairing. But, because the cylindrical portion of the Ares V faring is only 9.7 m tall (compared to 17.2 m in the center), it is not possible to have a fixed secondary mirror. Instead, a deployed tower is required.

## Segmented Mirror Telescope Options

A segmented-mirror space telescope is the only viable architecture for filled aperture sizes greater than 8 meters. We explored two designs, with apertures of 9.2 m and 16.8 m, which are derived from the *JWST* architecture.

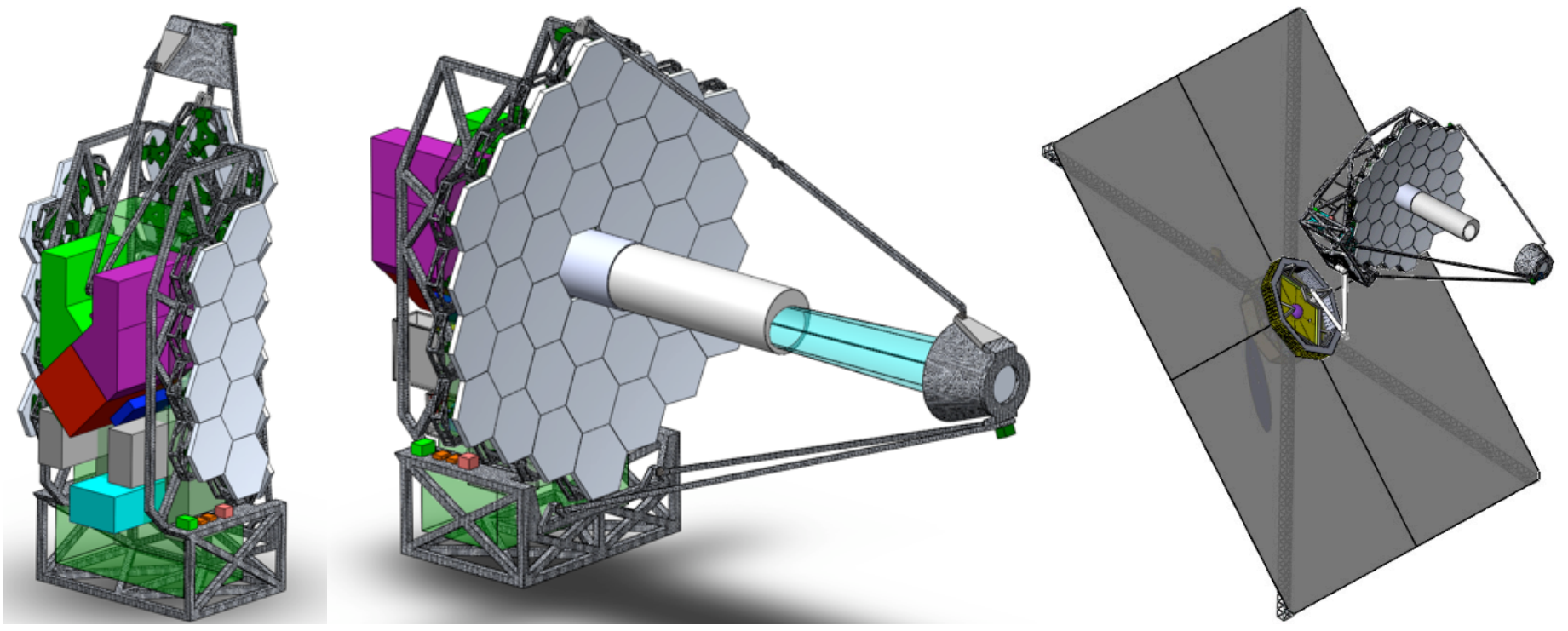

***Figure 6:*** (Left) Stowed 9.2-meter OTA. Colored boxes are instrument envelopes. (Center) 9.2-m *ATLAST* Deployed. (Right) Sunshield and arm-mounted OTA. Spacecraft bus is on sun side of sunshield.

***ATLAST 9.2-meter:*** The *ATLAST*-9.2m (see Figure 6) draws heavily on *JWST*, taking advantage of engineering and technologies already developed for that mission. The OTA has a segmented 9.2 m aperture, consisting of 36 hexagonal ultra low expansion (ULE) glass mirrors (1.315 m in size, flat-to-flat) on actuated hexapods. This technology was demonstrated early in the development of *JWST*. The OTA is thermally controlled using heaters on its backplane. The chord-folding of the primary mirror, similar to the *JWST* design, stows within the 6.5 m (outer diameter) fairing of an upgraded EELV. Upgraded EELVs with fairings of this size, coupled with lift capacity of more than 15,000 kg to SE-L2 are in the planning stages.

The TMA channel includes a wide-field camera, four combined fine-guidance sensors (FGS) and wavefront sensor (WFS) cameras, and accommodation for two additional science instruments (see Table 2). A continuous guide-star based wavefront sensing and control architecture that leverages the high TRL *JWST* technology is used in this design. Light from the telescope is fed to the FGS/WFS cameras for continuous guiding and wavefront sensing. A guide star selection mirror accesses a 4 × 4 arcminute FOV and steers an isolated, bright star onto the FGS and WFS detectors. The FGS produces centroids with 0.5 milli-arcsecond noise equivalent angle at 5 Hz. Two separated FGSs provide roll sensing. For wavefront sensing, the light is divided into two beams, and each beam is sent through a narrow band ($\Delta\lambda/\lambda \approx$ 1-5%) filter and a weak lens. One path uses a positive lens while the other uses a negative lens producing an out-of-focus image on either side of focus. To update the primary mirror positions and curvatures every five to twenty minutes, these images are processed onboard using phase retrieval. The secondary mirror position is adjusted as needed, using data obtained by three separated WFS cameras to break degeneracy with the primary mirror. A fourth FGS/WFS provides redundancy.

A thorough stray-light study concluded that a planar sunshield plus a central primary mirror baffle was sufficient to limit stray light to less than 10% of the zodiacal sky brightness while also eliminating "rogue path" light that would otherwise skirt around the secondary mirror. The deployed sunshield consists of four square-shaped layers of opaque Kapton film measuring 28 m on a side. To ease manufacturing, integration and test (I&T), and stowage procedures, each of the four layers is subdivided into four quadrants that are connected during I&T so that each deployed layer is continuous. Four 18-meter booms extend in a cruciform configuration to deploy the membranes.

The OTA is mounted on a multi-gimbal arm that provides OTA pitch motion, roll about the OTA line of sight, and center of mass trim for solar torque control while allowing the attitude of the spacecraft and sunshield to remain fixed relative to the Sun. This enables a very large field of regard with allowed pointing from 45 to 180° from the Sun. An active isolation system between the arm and OTA isolates spacecraft disturbances and, using the FGS sensor, provides a total image motion of ~1 milli-arcsec. The spacecraft bus provides coarse attitude control, propulsion, power, communication and data handling modeled after the *JWST* approach, but employs a modular architecture that allows servicing and reduces risk during I&T.

***ATLAST 16.8-meter:*** The OTA of *ATLAST-16* also consists of 36 hexagonal segments, in a 3-ring arrangement (see Fig. 7), but in this case each segment is ~2.4 m (flat-to-flat) producing an edge-to-edge primary-aperture diameter of 16.8 m. The mass of *ATLAST-16* can easily be kept within the Ares V lift capacity by using low areal density (15 kg/m$^2$) mirror material. One option is actuated hybrid mirrors [22]. AHMs consist of metallic nanolaminate facesheets that provide a highly reflective, optical quality surface on a SiC substrate with embedded actuators for figure control. The orientation of the segments is controlled in six degrees-of-freedom by rigid body actuators.

The segment rigid body and figure actuators are part of a wavefront sensing and control system required to maintain the telescope's optical performance. In addition, certain instruments will require a fine-steering mirror to stabilize images on a detector or spectrometer slit. The sensor for the segment rigid body actuators is an optical truss (*i.e.*, metrology system) that connects the segments to the secondary mirror. This truss provides a continuous update of the telescope configuration [23]. Multiple fine guidance sensors control the telescope body pointing.

The optical design is intimately connected with the strawman instrument suite (see Table 2). The *ATLAST-16m* wide FOV channel consists of two powered elements and a three-element stage that produces a compressed, collimated beam at a re-imaged pupil. The compressed beam has a manageable diameter of 30 cm.

Packing the 16.8-m telescope and its instrument suite is a significant challenge, even with the large volume in an Ares V fairing. Minimizing the primary-secondary mirror despace reduces mass, increases telescope stability, and increases the packaging options. To this end, we constrained the primary mirror focal ratio to 1.5 during the optical design phase. The geometry for the instrument volume depends on the folding scheme used for the telescope. To date, we have identified the following folding schemes for the telescope: chord-fold and hub-and-spoke.

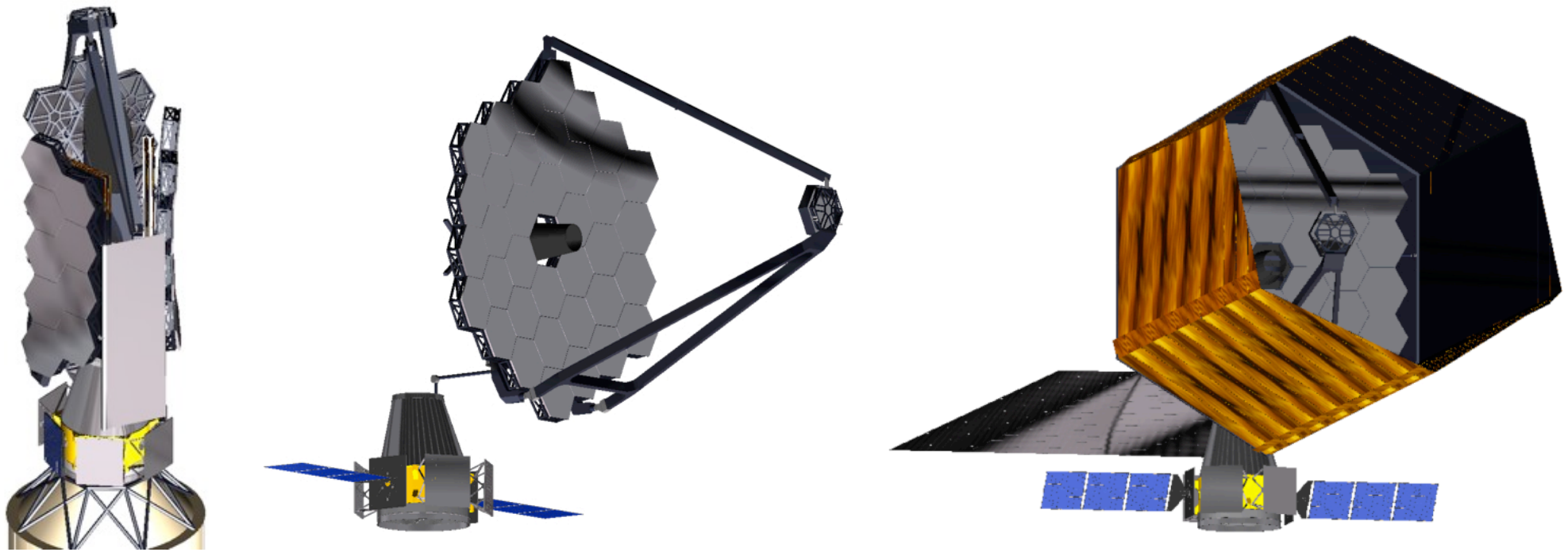

***Figure 7:*** (Left) Stowed 16.8-meter OTA. (Center) 16.8-m *ATLAST* deployed but the sunshade and "kite tail" (to mitigate solar torque) are not shown. (Right) 16.8-m shown with sunshield, "kite tail," and arm-mounted OTA.

The chord-fold design shares significant heritage with *JWST*. The primary mirror is folded to fit within the shroud and the secondary mirror is stowed along one edge. This architecture provides a cylindrical stowage volume 16 m in height x 8 m in width (see Fig. 7). Within this volume, a 16 m x 4.5 m diameter volume is available for instrument packaging. The disadvantages of this configuration are the increased center of gravity offset of the payload and reduced structural stiffness of the metering structure. In addition the secondary mirror must be replaced with a hexapod (vs. *JWST*'s tripod) to achieve reasonable (>10 Hz) decenter/despace frequencies.

The hub and spoke configuration folds the primary mirror around a 6.5 m diameter central hub, which supports petals on radial supports. The architecture provides a stowage volume 9 m in height x 8 m in width. The configuration provides a closer center of gravity offset of the payload and a simpler packaging for a conical sunshade. In addition, the structural stiffness of the primary mirror is greater and the secondary mirror support is simpler to package and deploy. The disadvantages of this configuration are the increased deployment complexity of the primary mirror and the reduced instrument packaging volume.

## 3. *ATLAST* Technology Drivers

The *ATLAST* technology development plan will bring the required technologies to readiness (Technology Readiness Level-TRL of six) as shown in Table 4. This section focuses on what we believe are the three most critical technology areas: *1)* large-aperture space optics capable of diffraction-limited performance at 500 nm, *2)* gigapixel focal plane arrays and high-efficiency UV detectors, and *3)* starlight suppression systems for exoplanet studies.

***Table 4: ATLAST* Technology Development Summary**

| Technology | Need | Needed Product to Achieve TRL6 | Current TRL | State of the Art |
|---|---|---|---|---|
| 8-m Monolithic Telescope | Launch loads, zero gravity figure | Mirror blank vibration test and gravity calibration | N/A | Gemini, VLT |
| Lightweight, lower-cost mirror segments | 1.3 to 2.5-m, < 15 kg/m$^2$, > 20 m$^2$/year, < \$1M/m$^2$ | Actuated Hybrid Mirror (AHM) or Glass Mirror at size, performance, and environments | 4 | JWST Architecture, AMSD Glass, AHM at 75 cm, Corrugated glass at 50 cm |
| Wavefront Sensing | Figure knowledge < 5 nm rms | Laser metrology and/or image based sensing | 4 | SIM-derived gauges, JWST Testbed Telescope |
| Segment Actuation | Resolution < 2nm | Actuator hexapod test and environments | 4 | JWST, Moog, PI |
| Visible Detectors | 8k x 8k arrays, and photon counting | Prototype performance and environments | 3-4 | E2V and TI photon counting CCDs |
| UV Detectors | Higher QE, 4 Mpix arrays | Prototype performance and environments | 4 | HST/ COS and STIS |
| UV Optics | UV IFS, coatings, dichroics | Prototype performance and environments | 4 | HST, FUSE |
| Internal Coronagraphs | $10^{-10}$ contrast over 20% passband | Prototype performance and environments | 4 | JPL HCIT tests |
| Starshade | Deployment and shape control | Sub-scale and partial shade tests | 3 | Beamline tests, deployment design with high TRL parts |

***Telescope Technology***: The diffraction-limited imaging at 500 nm that is needed for much of *ATLAST* science requires *HST*-quality mirror surface errors (5-10 nm rms) to meet the overall system wavefront error of 36 nm rms. For *ATLAST-8m*, solid meniscus monolithic glass, as demonstrated on ground-based telescopes, requires no new technology, but will require engineering to ensure survival of launch vibrations/acoustics and the proper gravity unloading. For the segmented *ATLAST* designs, hexagonal mirrors measuring 1.3 m and 2.4 m (flat-to-flat) are baselined for the 9.2-m and 16.8-m point designs, respectively. Some potential options for the mirror composition include the Advanced Mirror System Demonstrator (AMSD) ULE glass segments developed for *JWST*, Actuated Hybrid Mirrors (AHM), and corrugated glass technology [24]. Development goals for the segmented versions of *ATLAST* are primary mirror areal density of ~15 kg/m$^2$ and a systemic (including supporting structure and controls) areal density of ~30 kg/m$^2$. Some growth in areal density is acceptable as segment cost and production rate are as important as mass in choosing the mirror technology.

Modest development of the AMSD mirrors would be required to achieve better surface error polishing and gravity error removal. Both AHM and corrugated glass are replicated and show promise for faster and less expensive mirror production than the current processes. JPL, in conjunction with Lawrence Livermore National Laboratory and Northrop Grumman/Xinetics, have developed AHM over the last several years; see Figure 8. A metallic nanolaminate facesheet provides a high quality reflective surface, a SiC substrate provides structural support, and PMN surface parallel (in-plane) actuators provide surface figure control. Development is required to scale up segment size and improve surface figure. ITT has designed and tested corrugated glass mirror segments that promise faster and cheaper replicated production; however, they require development to demonstrate imaging quality, segment size and robustness in ULE glass. The TRL-6 milestone is a segment that meets environmental and wavefront error budget requirements.

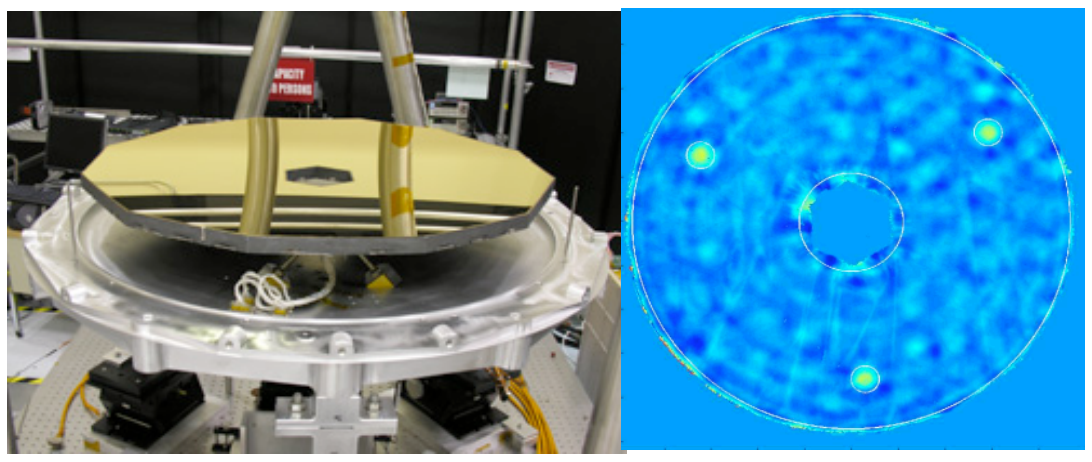

***Figure 8:*** Actuated hybrid mirrors (AHM) of less than 15 $kg/m^2$ have been tested and demonstrated 40 nm total WFE (gravity effect included), limited by high-spatial frequency mandrel polishing residual errors. With a higher quality mandrel, 10 nm surface figure error is achievable for meter-class segments.

Wavefront sensing and control (WFS&C) also needs to be advanced from *JWST* to meet visible image quality with an allocation of 10-15 nm rms wavefront error for WFS&C residual. The segment's positions must be measured to a few nanometers at a rate faster than their support structure's time constants. For *ATLAST-16m*, a JPL-developed laser metrology truss technology, derived from the SIM laser distance gauges, measures each of the segment's six degrees of freedom location (relative to the secondary mirror) at very high rates (up to 1 kHz); additional gauges locate downstream optics. The current TRL is 4; development is needed to advance the stability of the metrology and shrink the size and mass of the fiber-fed beam launchers. Periodic image-based WFS employs phase retrieval to correct drifts in the laser metrology. Actuators supporting the segments require a modest technology development to reduce the resolution to less than 2 nm and increase bandwidth. A TRL-6 demonstration of the telescope's full system operation is proposed with a subscale (6-m class), partially populated (three-segment) testbed meeting error budgets under full thermal, vacuum, vibration, and jitter environments. For more details on the relevant telescope technologies, see also the Astro2010 Technology Development paper by Werner et al. [25]

***Detector Technology:*** Gigapixel detector arrays for visible imaging and ~500 Megapixel arrays for NIR imaging are required for studies of resolved stellar populations, galaxy evolution, and structure formation. Such arrays can be built with existing technology. However, development would result in better science performance (lower noise), lower risk (less complex electronics), lower cost, and lower power consumption. The wide field cameras in all *ATLAST* designs are envisioned to have ~1 to 1.6 gigapixels per channel.

Exposure times invested in exoplanet and other faint object spectroscopy could be reduced by up to five times using photon counting detectors. Technology development of photon-counting CCDs is based on the low-light level CCDs built by E2V and the similar technology of Texas Instruments. The current TRL is 4; improvements in anti-reflection (AR) coatings, voltage swing, and charge induced clock noise would be demonstrated before the completion of a TRL-6 qualification program. In the longer term, CMOS based detectors are likely to be used. Detectors made by Fairchild, Sarnoff, and Fill Factory represent the state-of-the-art. Improvements in dark current, AR coating, in-pixel gain structure, and backside thinning would need to be demonstrated before the completion of a TRL-6 qualification program.

UV detectors have ample room for improvements in efficiency and format size. UV spectroscopy, in particular, requires coatings, optics and detectors that are highly efficient.

Current flight UV imaging detectors use CsI and CsTe photocathodes with 10% - 30% quantum efficiencies (QE). Photocathodes with QE of 30% - 80% using cesiated p-doped GaN have been produced in the lab, but have not yet been integrated into detector systems. Materials such as p-doped AlGaN and MgZnO should be developed for higher QE over a wider band, and better AR-coatings may be matched with p-channel radiation-hardened photon-counting CCDs. For the far UV, higher QEs also are required, and III/V materials are effective. All must be coupled with large-format intensifier and readout systems. EBCCDs and microchannel plate methods (ceramic and glass) compete for the highest QE and largest formats, and both should be pursued. The current TRL for high QE UV detectors is four; a downselect will precede TRL-6 qualification.

***Starlight Suppression Technology:*** An internal coronagraph or external occulter than can suppress starlight by a factor of $10^{10}$ is required to characterize Earth-sized exoplanets. The best starlight suppression technology is not yet obvious. Fortunately, there are multiple options. The technology development plan addresses the viability of 1) a visible nulling coronagraph (VNC) that any telescope could use (development costs might be shared with large ground-based telescopes); 2) a Lyot-type coronagraph (for use with monolithic telescopes); and 3) an external starshade (a separate spacecraft creating a star shadow at the telescope). All three methods are currently at or below TRL-4 [26]. Recognizing the importance of *ATLAST* to exoplanet science, we would fund development of all three methods for starlight suppression, with a downselect 3-5 years before the scheduled TRL-6 milestone.

There are a number of coronagraphic configurations, each having tradeoffs between achievable contrast, IWA, throughput, aberration sensitivity, and ease-of-fabrication. Most techniques do not work on conventional (4-arm spider) obscured or segmented systems, but some non-conventional spider configurations (linear support) appear to allow performance competitive with off-axis systems. The VNC approach is probably the only viable solution for *internal* suppression with a segmented telescope. VNC development is required to demonstrate $10^{-10}$ contrast over a 23% passband. This requires the development of a spatial filter array (1027 fiber bundle), deformable mirror (MEMS 1027 segment), and an achromatic phase shifter. Testbeds at JPL and GSFC have presently achieved $10^{-7}$ contrast ratios. Coronagraphic technology development is critically dependent on advancements in high-precision, high-density deformable mirrors (>96 actuators across pupil and sub-Angstom stroke resolution).

Starshade technology development is well described in the New Worlds Observer RFI response. Starshade theoretical performance has been validated by at least 4 independent algorithms and, in the lab, by two beamline testbeds. Detailed CAD models exist for the *NWO* 50-m starshade, which use high TRL-components (membranes, hinges, latches, booms). For *ATLAST*, larger starshades are required: 80 m and 90 m for our 8-m and 16.8-m telescopes, respectively. The key challenges are primarily deployment reliability and shape control. *ATLAST* starshade separations of 165,000 km to 185,000 km, while large, do not present any challenging formation flying or orbital dynamics issues but put additional requirements on the starshade propulsion system. The starshade technology developments are addressed through increasingly larger subscale models with TRL-6 being demonstrated through beamline tests, a half-scale quarter-section deployment, and a full-scale single petal deployment, performance, and environmental testing. The astrometric sensor and NASA's Evolutionary Xenon Thruster (NEXT) ion engine needed to align the starshade already will be TRL-6 or greater by other projects: USNO's JMAPS and NASA's in-space propulsion program.

## 4. *ATLAST* Technology Development Schedule

The *ATLAST* Technology Development Plan is summarized in Figure 9. This roadmap shows the TRL 5 and 6 milestones, downselect logic, funding profile, and interactions within the technology development tasks. Assuming a 2011 start for technology development, the project would initially begin funding high priority technology developments at a low level. The schedule is divided in to six periods. If we can afford a "full speed ahead" pace, each period would last one year and all technologies would reach TRL-6 in 2017. A slower schedule of 1.5 years per period would be completed by 2020. Milestones are placed at the TRL-5 level and coincide with downselects in three areas (mirror segment technology, visible detector technology, and starlight suppression technology). All technologies would plan to complete their TRL-6 milestones with at least a year margin before a mission PDR (NASA requires TRL-6 at PDR), allowing a healthy schedule margin of two months per year. A parallel pre-formulation mission study would inform the technology trades and downselects. Downselect decisions on the launch vehicle and primary mirror architecture (monolith versus segmented) are made by the end of period three.

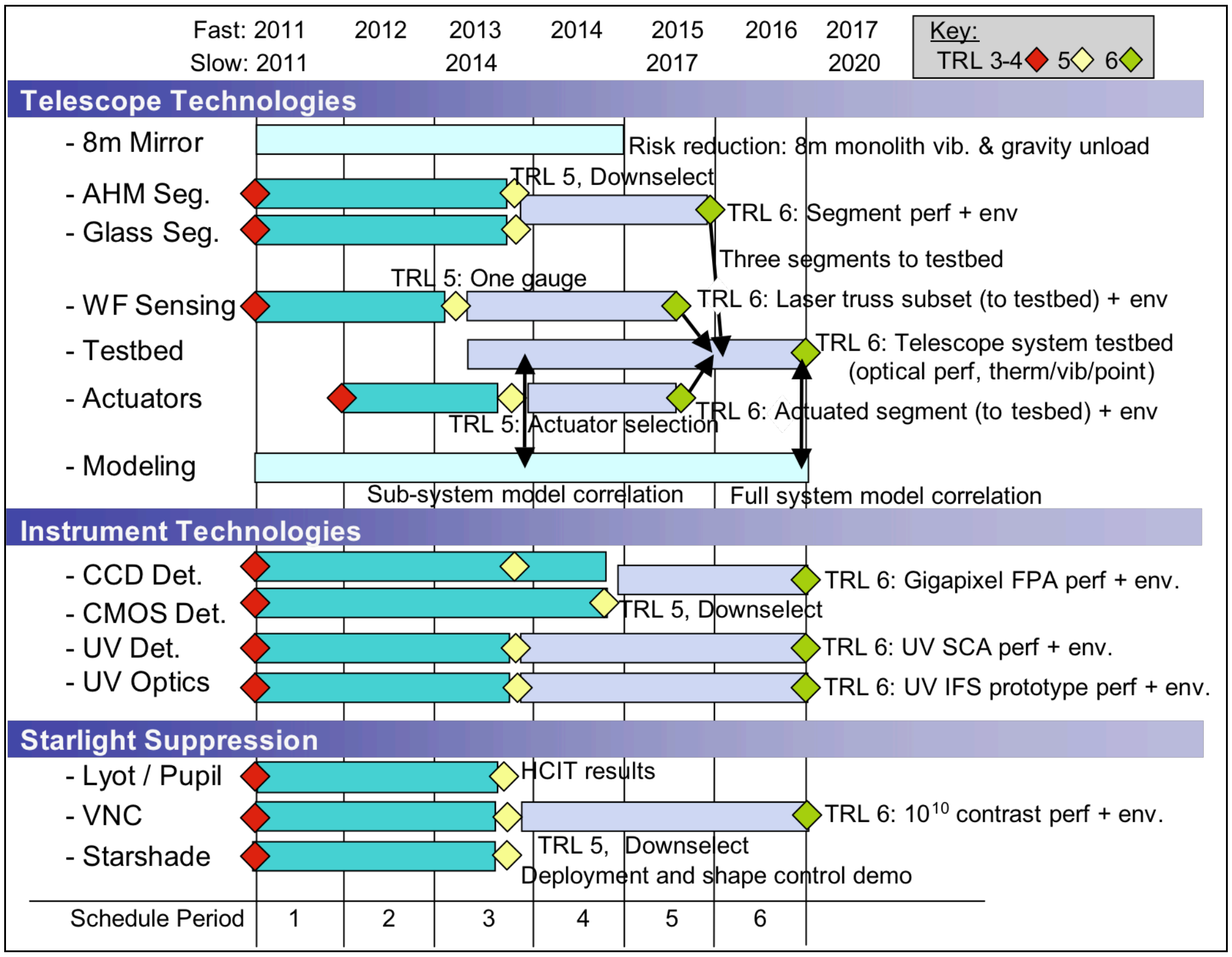

***Figure 9:*** Proposed ATLAST Technology Development Plan & Schedule

The telescope technology roadmap includes funding for AHM and corrugated-glass segments through a TRL-5 downselect with AMSD segments as an offramp. One full size segment is then taken through TRL-6 qualification and three 1.3-m segments are delivered to the testbed task for system validation. Wavefront sensing with laser metrology is matured to TRL-6 and a set of gauges delivered to the testbed; image based sensing and better thermal stability is an offramp. Actuators already are close to meeting their performance and flight qualification requirements so they only require modest funding to a TRL-6 demonstration of a hexapod-actuated segment; a set of actuators is delivered to the testbed. The testbed demonstrates the end-to-end performance thereby qualifying the full telescope system at TRL-6. Integrated modeling is a critical element of proving technology at full scale in space and is based on sub-scale ground testing. This effort iterates with other tasks to analyze the designs and shape the testing procedures so that uncertain elements of the model are tested, thus providing a "two-way street." Technology investment in low-cost lightweight mirror benefits not only a large space telescope but also smaller missions; a single segment could serve a SMEX-class telescope.

The instruments section of the roadmap addresses detectors and some instrument technologies for UV optics (filters, gratings, and coatings). Advancements in all of these areas will improve SNR in a given exposure time or significantly reduce exposure time required to reach a given SNR. Investments on total throughput efficiency may be traded against aperture size, provided spatial resolutions are not degraded below the science requirements. The technology investments in detectors would again serve both *ATLAST* and smaller class missions.

The starlight suppression technology is advanced in all three methods through TRL-5 since the *ATLAST* optical telescope assembly architecture is still open. A strict downselect may not be warranted, as *ATLAST* is compatible with both an internal method and a starshade. If a separate exoplanet probe mission were selected to precede *ATLAST*, both could potentially share technology development costs, thus reducing *ATLAST's* overall costs.

Our explicit cost methodology and estimates are given in the next section.

**Summary by Point Design:**

***ATLAST-8m:*** Because of the high-TRL of 8-m mirrors, no mirror technology development is required. Rather, we strongly suggest elimination of schedule risk by determining the mirror blank's ability to survive launch before Phase A. If the existing Zerodur blank is not suitable then a long-lead replacement blank procurement process could be initiated. Flight qualification for an 8-m primary mirror is accomplished by scaling-up and qualifying standard engineering methods proven on ground telescopes, *HST* and *Kepler*. Polishing all surfaces of the blank raises its stress design limit to 7000 psi and allows inspection of any internal defects which could limit its strength. Actual blank strength can be characterized via proof and sine-burst tests. Simultaneously, it is possible to mitigate on-orbit performance risk by absolutely calibrating the blank's g-release shape change.

***ATLAST-9.2m:*** Technology development would be moderately cheaper (vs. *ATLAST-16m*) for the telescope technology portion of the roadmap due to the higher reliance on existing AMSD segment and image-based wavefront sensing technologies.

***ATLAST-16m:*** This concept requires the full development investment because novel WFS&C systems and multiple mirror material and fabrication technologies need to be explored.

The major milestones of the technology development effort are the TRL-5 and 6 gates. A technology evaluation board made up of non-advocate experts in each of the areas would be formed for the duration of the effort. The board's first job would be to outline the exact demonstration that would satisfy the TRL-gates. Then, as the development efforts matured and completed these demonstrations, the test data and modeling would be presented back to the board for TRL-gate review. The board's final review is a Technology Non-Advocate Review that would validate all required mission technologies at TRL-6 or higher, thereby supporting a successful PDR and allowing the mission to proceed to implementation. Once technology development is complete and the key decision point given, the elapsed time between the start of phase A and launch is 8 to 12 years, depending on the design selected.

## 5. *ATLAST* Technology Development Cost Estimate

The cost estimates for the *ATLAST* Technology Development Plan (TDP) are given in Table 5. Experts in the 14 technology areas generated all our TDP costs. These experts, representing NASA/GSFC, NASA/MSFC, JPL, Ball Aerospace, Northrop Grumman, Xinetics, and ITT, have planned tasks to meet TRL-6.

***Table 5:*** **Technology Development Cost Estimates (FY09 $M, No Reserves)**

| **Period:** | **1** | **2** | **3** | **4** | **5** | **6** | | **Totals** |
|---|---|---|---|---|---|---|---|---|
| **Fast:** | **2011** | **2012** | **2013** | **2014** | **2015** | **2016** | **2017** | |
| **Slow:** | **2011** | **2013** | **2014** | **2016** | **2017** | **2019** | **2020** | |
| **8-m Telescope** | | | | | | | | 20.0 |
| - Blank Vib. Test | 5.0 | 5.0 | 5.0 | 5.0 | | | | 20.0 |
| **Segmented Telescope** | | | | | | | | 147.8 |
| - AHM Seg. | 4.6 | 5.8 | 6.4 | 9.3 | 12.9 | | | 39.0 |
| - Glass Seg. | 2.2 | 3.0 | 3.8 | 11.0 | 20.0 | | | 40.0 |
| - [downselect savings] | | | | -11.3 | -20.9 | | | -32.2 |
| - Laser Truss | 1.0 | 2.0 | 5.0 | 5.0 | 5.0 | | | 18.0 |
| - Testbed | | | 5.0 | 20.0 | 20.0 | 20.0 | | 65.0 |
| - Actuators | | 0.5 | 1.5 | 2.5 | 2.5 | | | 7.0 |
| - Modeling | 1.0 | 1.5 | 1.5 | 2.0 | 2.0 | 3.0 | | 11.0 |
| **Instruments** | | | | | | | | 60.1 |
| - CCD Detectors | 1.6 | 0.6 | 1.0 | 1.0 | 4.0 | 4.0 | | 12.2 |
| - CMOS Det. | 0.9 | 1.2 | 0.9 | 0.9 | 4.0 | 4.0 | | 11.9 |
| - [downselect savings] | | | | | -4.0 | -4.0 | | -8.0 |
| - UV Detectors | 3.0 | 5.0 | 6.4 | 5.3 | 5.5 | 2.7 | | 27.9 |
| - UV Optics | 0.9 | 3.1 | 4.3 | 3.5 | 2.8 | 1.6 | | 16.1 |
| **Starlight Suppression System** | | | | | | | | 59.5 |
| - Lyot / Pupil Cor. | 6.3 | 7.7 | 5.6 | 6.0 | 5.5 | 4.0 | | 35.1 |
| - VNC | 0.9 | 1.7 | 4.4 | 2.5 | 16.9 | 5.2 | | 31.5 |
| - Starshade | 1.5 | 1.5 | 5.0 | 5.0 | 10.5 | 10.5 | | 34.0 |
| - [downselect savings] | | | | -5.5 | -20.9 | -14.7 | | -41.1 |
| **Technology Development Sub-Total:** | | | | | | | | 287.4 |
| **Mission Study** | | | | | | | | 48.0 |
| - Science WG & Studies | 4.0 | 4.0 | 4.0 | 4.0 | 4.0 | 4.0 | | 24.0 |
| - Project Office & Eng. Studies | 4.0 | 4.0 | 4.0 | 4.0 | 4.0 | 4.0 | | 24.0 |
| **Totals** | 36.9 | 46.5 | 63.8 | 70.2 | 73.8 | 44.3 | | 335.4 |

The tasks in our TDP are spread over six schedule periods. Downselects in three areas after period three allow higher funding on the most promising technology for the final three periods. Informing this downselect schedule requires mission decisions about launch vehicle and segmented versus monolithic telescope. All estimated costs are in constant FY2009 dollars and do not include reserves. The time phasing of the technology development cost grows nearly linearly from $29M in the first period to $66M in the fifth period. The segmented telescope technology investment area is the greatest, at $148M. This value is consistent with the sum of cost estimates for lightweight large aperture and WFS&C given by the NASA Advanced Planning and Integration Office roadmap activity in 2005. That effort included members from NASA, industry, and the National Reconnaissance Office. The detector development benefits any space UVOIR instrument regardless of mission size. The starlight suppression technology builds, in part, on the investment in *TPF-C* at JPL over the last decade. The *ATLAST* TDP costs here assume full coverage of the VNC and starshade costs. We assume some of the Lyot/pupil internal coronagraph technology development costs will be shared with a smaller exoplanet probe mission; only 70% of the grass roots cost estimate is included for the first three periods; 100% in the last 3, post-downselect.

A total *ATLAST* technology development project would cost $287M if all downselects were kept open until the TRL-5 milestones in period three. This does not include reserves, which would be held at the program level. The range of technology development cost estimates for *ATLAST-8m* spans $55M - $149M depending on the degree to which early downselects are taken. The cost estimate range for the *ATLAST* segmented-mirror concepts spans $115M - $267M, again, depending on the timeline for early downselects taken.

In addition to the cost estimates for technology development given above, there would be a parallel mission study integrated with the technology development work. This was roughly estimated at $8M per period, with half supporting an *ATLAST* science working group and competitively awarded grants and half supporting a pre-formulation project team and competitively awarded engineering studies.

### *ATLAST* Mission Life Cycle Cost Estimates

Mission costs and parameters for *ATLAST-8m* and *ATLAST-9.2m* have been studied at MSFC and GSFC, respectively.

The *ATLAST-9.2m* mission concept was studied in the Integrated Design Center (IDC) at Goddard Space Flight Center for three weeks during February and March 2009. A combination of parametric and "grassroots" estimates of mission cost were developed for the life cycle of the mission in **constant FY08 dollars**. The total cost for phases A through E, including 30% reserve on all mission elements, is **$5.6B**. This total includes 5 years of mission operation costs ($200M) plus $125M to the science community for research and analysis of the data. The total does not include the cost of the launch vehicle, which is not currently known; the IDC provided an approximate cost of $400M for this item. The *ATLAST-9.2m* cost by mission phase is given in Table 6.

***Table 6***

| ***ATLAST-9.2m* (FY08 $)** | |
|---|---|
| Phase A | 340M |
| Phase B | 1,400M |
| Phase C/D | 3,470M |
| Phase E | 375M |

The *ATLAST-8m* cost was studied at the Advanced Concepts Research Facility at Marshall Space Flight Center during the February – March 2009 timeframe. The result of this study was a cost estimate for the *ATLAST-8m* payload. The MSFC estimate is being reviewed for completeness and compared with the GSFC IDC cost methodologies to finalize a life cycle cost. The preliminary estimates indicate that the full life cycle cost (including a 5-year phase E period) for *ATLAST-8m* is \$4.5B to \$5B. However, final cost estimates were not available at the time this RFI response was submitted. They are expected prior to June.

The above life cycle costs do not include the cost of an external occulter and its spacecraft. The above costs, however, do include the estimated cost of an internal coronagraphic instrument.

A detailed report on the *ATLAST* technology development plan and mission concepts will be submitted to NASA on April 24 as the prime deliverable of our Astrophysics Strategic Mission Concept Study. A life cycle cost estimate for *ATLAST-16m* has not been performed.

## Acknowledgements

Goddard Space Flight Center managed the NASA Astrophysics Strategic Mission Concept Study of *ATLAST*, which formed the basis of this white paper. Portions of the work were performed at the Space Telescope Science Institute, Ball Aerospace & Technologies Corp., Goddard Space Flight Center, Marshall Space Flight Center, the Jet Propulsion Laboratory, the California Institute of Technology, Northrop Grumman Aerospace Systems, the University of Colorado at Boulder, the University of Massachusetts at Amherst, and Princeton University.

## Table of Acronym Definitions

| | | | | | |
|---|---|---|---|---|---|
| AHM | Actuated Hybrid Mirror | IDC | Integrated Design Center | SMEX | Small Explorer Mission |
| AMSD | Advanced Mirror System Demonstrator | IGM | Intergalactic Medium | SNR | Signal-to-Noise Ratio |
| ATLAST | Advanced Technology Large Aperture Space Telescope | IMF | Initial Mass Function | TDP | Technology Development Plan |
| CAD | Computer-Aided Design | IWA | Inner Working Angle | TMT | Thirty Meter Telescope |
| CCD | Charge Coupled Device | JMAPS | Joint Milli-Arcsecond Pathfinder Survey | TPF-C | Terrestrial Planet Finder Coronagraph Mission |
| CMOS | Complementary Metal–Oxide Semiconductor | JPL | Jet Propulsion Laboratory | TRL | Technology Readiness Level |
| COS | Cosmic Origins Spectrograph | JWST | James Webb Space Telescope | ULE | Ultra Low Expansion |
| CTE | Coefficient of Thermal Expansion | MEMS | Micro Electro-Mechanical Systems | UV | Ultraviolet |
| EELV | Enhanced Expendable Launch Vehicle | MSFC | Marshall Space Flight Center | UVOIR | The wavelength range from ~110 to ~2500 nm |
| FGS | Fine Guidance Sensor | NASA | National Aeronautics and Space Administration | VLT | Very Large Telescope |
| FOV | Field of View | NEXT | NASA's Evolutionary Xenon Thruster | VNC | Visible Nulling Coronagraph |
| FUSE | Far Ultraviolet Spectroscopic Explorer | NWO | New Worlds Observer | WFE | Wavefront Error |
| GALEX | Galaxy Evolution Explorer | OTA | Optical Telescope Assembly | WFS | Wavefront Sensing |
| GSFC | Goddard Space Flight Center | PDR | Preliminary Design Review | WFS&C | Wavefront Sensing and Control |
| HST | Hubble Space Telescope | SE-L2 | Sun-Earth $2^{nd}$ Lagrange point | | |
| HZ | Habitable Zone | SIM | Space Interferometry Mission | | |